\documentclass{article}
\usepackage{epsfig}
\title{\bf Bulk Viscosity of Magnetized Neutron Star Matter}

\author{J D Anand, % \thanks {E--mail:jda@ducos.ernet.in},
V K Gupta, Ashok Goyal, S Singh, Kanupriya Goswami \\
{\em  Department of Physics and Astrophysics,}\\
{\em  University of Delhi,Delhi-110007,India.}\\
{\em  Inter University Centre for Astronomy and Astrophysics}\\
{\em  Ganeshkhind,Pune-411007,India.}\\
}

\begin{document}

\maketitle
%\begin{center}
%{\em  Department of Physics and Astrophysics,}\\
%{\em  University of Delhi,Delhi-110007,India.}\\
%{\em  Inter University Centre for Astronomy and Astrophysics}\\
%{\em  Ganeshkhind,Pune-411007,India.}\\
%\end{center}

\begin{abstract}
\noindent We study the effect of magnetic field on the bulk viscosity of 
nuclear matter in neutron stars. We employ the framework of relativistic
mean field theory to observe the dense nuclear matter in neutron stars.
The effects are first studied for the case when the magnetic field does 
not exceed the critical value to confine the electrons to the lowest 
Landau levels. We then consider the case of intense magnetic field to 
evaluate viscosity for the URCA processes and 
show that the inequality $p_{F}(e)+p_{F}(p)\geq p_{F}(n)$ is no 
longer required to be satisfied for the URCA processes  to proceed. 
\end{abstract}
\pagebreak

\begin{section} {Introduction}
\indent After the discovery of magnetars there has been a lot of interest 
in the study of the effects of strong magnetic fields on the various 
astrophysical phenomena. The observed soft gamma repeaters discovered 
recently by Kuoveliotou C et al [1] and Kulkarni and Thomson 
[2] show the presence of strong magnetic fields of upto 
$10^{15}$ Gauss. Though the core of neutron stars is poorly understood, 
magnetic fields as strong as $B\sim 10^{18}$ Gauss may be assumed in the 
core by the scalar virial theorem.
It has been argued that even higher fields ( $B\sim 10^{20}$ Gauss) could be 
consistent with the virial theorem [3]. It is therefore 
important to investigate the global properties of these objects, particularly 
a detailed investigation of the stability of dense stellar matter in the 
presence of ultra strong magnetic fields. This requires a study of the 
bulk viscosity of such objects, since the damping of vibrations and secular 
instability of rapidly rotating neutron stars depend upon the bulk viscosity.
The bulk viscosity of neutron stars arises mainly due to deviation from 
$\beta$ equilibrium and the ensuing non-equilibrium reactions.
These important reactions are so called URCA processes:

\vspace{0.5cm}
\begin{equation}
n\longrightarrow p+e^{-}+\bar{\nu_{e}} 
\end{equation}

\begin{equation}  
p+e^{-}\longrightarrow n+\nu_{e}
\end{equation}
 
and the modified URCA processes:

\begin{equation}		
n+n\rightarrow n+p+e^{-}+\bar{\nu_{e}}
\end{equation}

\begin{equation}	
n+p+e^{-}\rightarrow n+n+\nu_{e}
\end{equation}

At low temperatures for degenerate nuclear matter, the direct URCA
process can take place only near the fermi energies of participating 
particles. Simultaneous conservation of energy and momentum puts the constraint

\begin{equation} 
p_{F}(e)+p_{F}(p)\geq p_{F}(n)
\end{equation}

which must be satisfied for the URCA processes to take place. In most 
models of nuclear matter this constraint 
is satisfied only at high densities,viz, $n_{b} \geq 1.5~ n_{0}$, where $n_{0}=0.16~ fm^{-3}$,is the nuclear saturation density.
There have been many studies of the bulk viscosity of neutron stars 
under the usual situation of zero magnetic field [4-8].
In this note we report the calculations of bulk viscosity of neutron stars 
from the URCA processes in the presence of magnetic field. Such a study is 
of interest because significant changes in the equation of state of a 
magnetized star could lead to similar changes in the bulk viscosity as well.
We have considered a wide range of magnetic fields in our study, from 
``low'' to ``very high''magnetic fields ($\geq 10^{19} G$). As has been 
discussed above, such extremely strong magnetic fields in the interior of
 stellar objects can not be completely discounted  and have infact been 
already studied. For the description of nuclear matter at high densities 
which are found in the interior of neutron stars, we use the frame work of 
relativistic nuclear mean field theory, including the $\rho$ contribution, as
extended to include strong interactions among the particle species involved.
As is well known, this would give a fairly good description of strong  
interactions among the various particles and can be easily extended to 
include higher resonances as well.
In Section 2 we derive expressions for bulk viscosity both in the 
high and low magnetic field limits. In Section 3 we present our results 
followed by a brief discussions.

\end{section}

\begin{section}{Derivation of Bulk Viscosity}

\noindent In this Section we derive expressions for the bulk viscosity
in the presence of magnetic field. The presence of magnetic field affects the
bulk viscosity only through the changes in the reaction rates as well
as through the modification of the chemical composition of neutron
star matter. We therefore adopt an approach similar to that of 
Wang and Lu {\cite {label 4}}, Sawyer
{\cite{label 5}}, Madsen{\cite {label 6}} and Gupta et al
{\cite {label 7}} for the zero magnetic field case.
The bulk viscosity $\zeta$ is defined by

\begin{equation}
\zeta=2(\frac{dW}{dt})_{av}\frac{1}{v_{0}}(\frac{v_{0}}{\Delta v})^{2}
	\frac{1}{\omega^{2}}
\end{equation}
										
Here $v_{0}$ is the specific volume of the star in equilibrium 
configuration, ${\mit\Delta v}$ is the amplitude of the periodic 
perturbation with period $\tau=\frac{2\pi}{\omega}$:
\begin{equation}
v(t)=v_{0}+{\mit\Delta v}\sin(\frac{2\pi t}{\tau})
\end{equation}

The quantity $(\frac{dW}{dt})_{av}$ is the mean dissipation rate of 
energy per unit mass and is given by the equation
\begin{equation}
(\frac{dW}{dt})_{av}=-\frac{1}{\tau}\int P(t)\frac{dv}{dt}dt
\end{equation}

In the above the pressure $P(t)$ can be expressed near its equilibrium 
value $P_{0}$, as
\begin{eqnarray}
P(t)=P_{0}+(\frac{\partial P}{\partial v})_{0}\delta v+
	(\frac{\partial P}{\partial n_{p}})_{0}\delta n_{p} +
	(\frac{\partial P}{\partial n_{e}})_{0}\delta n_{e}+	
	(\frac{\partial P}{\partial n_{n}})_{0}\delta n_{n}
\end{eqnarray}

The change in the number of neutrons, protons and electrons per unit mass 
over a time interval $(0,t)$ due to the URCA reactions (1) and (2) is given by
\begin{equation}
-\delta n_{n}=\delta n_{p}=\delta n_{e}=\int_{0}^{t}\frac{dn_{p}}{dt}dt
\end{equation}

The net rate of production of protons, $\frac{dn_{p}}{dt}$ is given by the 
difference between the rates ${\mit\Gamma_{1}}$ and ${\mit\Gamma_{2}}$
of reactions (1) and (2) respectively. At equilibrium the two rates
are obviously equal and the chemical potentials satisfy the equality

\begin{equation}
\Delta\mu=\mu_{n}-\mu_{p}-\mu_{e}=0
\end{equation}

A small volume perturbation brings about a small change in the chemical 
potentials and the above inequality is no longer satisfied, now  
${\mit\Delta\mu}$ is not  zero and consequently  the reaction rates of 
the processes (1) and (2) 
are no longer equal. The net rate of production of protons will thus 
depend upon the value of ${\mit\Delta\mu}$.
In the linear approximation, $\frac{\Delta\mu}{kT}\ll 1$, the net rate 
is proportional to $\Delta\mu$ and thus can be written as

\begin{equation}
\frac{dn_{p}}{dt}={\mit\Gamma_{1}}-{\mit\Gamma_{2}}=-\lambda{\mit\Delta\mu}
\end{equation}
 
Using the thermodynamic relation
\begin{equation}
\frac{\partial P}{\partial n_{i}}=-\frac{\partial \mu_{i}}{\partial v}
\end{equation}
in equation (9) and employing equation (10), we obtain
\begin{eqnarray}
%\begin{split}
\delta P 
&=& P(t)-P_{0}\nonumber	\\
&&{}	 = -\frac{\partial(\delta\mu)}{\partial v}
		\int_{0}^{t}\lambda\delta\mu(t)dt	
%\end{split}
\end{eqnarray}

The change in the chemical potential $\delta\mu(t)$ arises due to 
a change in the specific volume $\delta v$ and changes in the 
concentrations of various species, viz, neutrons, protons and electrons. Thus

\begin{eqnarray}
\delta\mu(t)
&=& \delta\mu(0)+(\frac{\partial\delta\mu}{\partial v})_{0}\delta v
	+(\frac{\partial\delta\mu}{\partial n_{n}})_{0}\delta n_{n}	
		+(\frac{\partial\delta\mu}{\partial n_{p}})_{0}\delta n_{p}
		+(\frac{\partial\delta\mu}{\partial n_{e}})_{0}\delta n_{e}
\end{eqnarray}

Using equations (10)-(12), we arrive at the following equations for 
$\delta\mu$:
\begin{equation}
\frac{d\delta\mu}{dt}=\omega A \frac{{\mit\Delta v}}{v_{0}}\cos(\omega t)-
			C\lambda\delta\mu
\end{equation}
where
\begin{equation}
A=v_{0}(\frac{\partial\delta\mu}{\partial v})_{0}
\end{equation}

\begin{equation}
C=v_{0}(\frac{\partial\delta\mu}{\partial n_{p}}+
	\frac{\partial\delta\mu}{\partial n_{e}}-
	\frac{\partial\delta\mu}{\partial n_{n}})_{0}
\end{equation}

Since for small perturbations, $\lambda$, A and C are constants, equatioin 
(15)
can be solved analytically to give
\begin{equation}
\delta\mu=\frac{\omega A}{\omega^{2}+C^{2}\lambda^{2}}
	\frac{{\mit\Delta v}}{v_{0}}
	[-C\lambda e^{-C\lambda t}+\omega\sin(\omega t)
	+C\lambda\cos(\omega t)]
\end{equation}

Finally using equations (6), (8), (10), (11), (14) and (17)-(19), we obtain
the following expressions for $\zeta$
\begin{eqnarray}
\zeta=\frac{A^{2}\lambda}{\omega^{2}+C^{2}\lambda^{2}}
	[1-\frac{\omega C\lambda}{\pi}
	\frac{1-e^{-C\lambda\tau}}{\omega^{2}+C^{2}\lambda^{2}}]
\end{eqnarray}

Given the number densities of the these particle species in terms  of
their respective chemical potentials, one can determine the coefficients 
A and C; given the rates  $\Gamma_{1}$ and $\Gamma_{2}$
for the two processes (1) and (2) one can determine $\lambda$ and 
hence $\zeta$ for any given baryon density and temperature.
For the description of the dense nuclear matter we have used Walecka's 
mean field theory [8-10]. In the mean field approximation we allow the mesons fields $\sigma$, $\omega$ and $\rho$ to acquire density dependent average 
values. From the effective langrangian one can then read 
off the effective masses and chemical potentials $\bar m_{i}$ and 
$\bar\mu_{i}$. The fermi momenta, are in turn related to these effective 
quantities by
\begin{equation}
\bar k_{i}=\sqrt{\bar\mu_{i}^{2}-\bar m_{i}^{2}}
\end{equation}

The various chemical potentials and masses required in these expressions 
are obtained from the self consistent solution of the mean field equations:
\begin{eqnarray}
m_{\sigma}^{2}\bar\sigma+bm_{n}b_{\sigma N}^{3}\bar\sigma^{2}+
	cg_{\sigma N}^{4}\bar\sigma^{3}-g_{\sigma N}
	(n_{p}^{s}+n_{n}^{s})=0\nonumber \\
m_{\omega}^{2}\bar\omega_{0}=g_{\omega N}(n_{p}+n_{n})\nonumber  \\
m_{\rho}^{2}\bar\rho_{0}=\frac{1}{2}g_{\rho N}(n_{p}-n_{n})\nonumber  \\
m_{n}-\bar m_{n}=m_{p}-\bar m_{p}=g_{\sigma n}\bar\sigma_{0}\nonumber  \\
\bar\mu_{n}=\mu_{n}-g_{\omega N}\bar\omega_{0}+
		\frac{1}{2}g_{\rho N}\bar\rho_{0}\nonumber  \\
\bar\mu_{p}=\mu_{p}-g_{\omega N}\bar\omega_{0}-
		\frac{1}{2}g{\rho N}\bar\rho_{0}\nonumber  \\
\end{eqnarray}
The number  densities $n_{i}$ and the scalar densities $n_{i}^{s}$ 
{\cite{label 9}} are given in terms of these quantities by the expressions

\begin{equation}
n_{i}=2\int \frac{d^{3}p}{(2\pi)^{3}}
	({1+e^{\frac{E_{i}-\bar\mu_{i}}{T}}})^{-1}
\end{equation}

\begin{equation}
n_{i}^{s}=2 \bar m_{i}\int \frac{d^{3}p}{(2\pi)^{3}}\frac{1}{E_{i}}
		({1+e^{\frac{E_{i}-\bar\mu_{i}}{T}}})^{-1}
\end{equation}
 
where 
\begin{equation}
E_{i}=\sqrt{p^{2}+\bar m_{i}^{2}}
\end{equation}

Since the chemical potentials $\mu_{n}$ and $\mu_{p}$ are of the order of a few
hundred MeV, even for temperatures  upto a few MeV the matter is completely 
degenerate, this leads to the following expressions for $n_{i}$ and 
$n_{i}^{s}$ :
\begin{equation}
n_{i}=\frac{1}{3\pi^{2}}(\bar\mu_{i}-\bar m_{i}^{2})^{\frac{3}{2}}
\end{equation}
\begin{equation}
n_{i}^{s}=\frac{\bar m_{i}}{2\pi^{2}}\bar\mu_{i}
		(\bar\mu_{i}^{2}-\bar m_{i}^{2})^{\frac{1}{2}}
\end{equation}
where i=n, p. To evaluate $n_{e}$, set $\bar\mu_{e}=\mu_{e}$ and 
$\bar m_{e}=m_{e}$ in equation(26).

It now remains to introduce the magnetic field. In the presence of 
 constant magnetic field B in the z-direction, the energy of a charged 
particle of mass m and charge e is given by
\begin{equation}
E^{2}=p_{z}^{2}+m^{2}+2\nu eB
\end{equation}

where the quantum number $\nu$ is given by
\begin{equation}
\nu=\nu_{L}+\frac{1}{2}+\sigma
\end{equation}
for the Landau level $\nu_{L}=0,1,2....$ and spin $\sigma=\pm \frac{1}{2}$.

In the presence of a magnetic field, the number densities of electrons and 
protons are given by
\begin{equation}
n_{e}=\frac{eBv}{2\pi^{2}}\sum_{\nu=0}^{\nu_{max}}(2-\delta_{\nu,0})
	\sqrt{\mu_{e}^{2}-m_{e}^{2}-2\nu eB}
\end{equation}	

and
\begin{equation}
n_{p}=\frac{eBv}{2\pi^{2}}\sum_{\nu=0}^{\nu_{max}}(2-\delta_{\nu,0})
	\sqrt{\bar\mu_{p}^{2}-\bar m_{p}^{2}-2\nu eB} 
\end{equation}

whereas the  neutron number density expression remains  
unchanged namely
\begin{equation}
n_{n}=\frac{1}{3\pi^{2}}(\bar\mu_{n}^{2}-\bar m_{n}^{2})^{\frac{3}{2}}
\end{equation}
 In the decay rates the usual sum over the charged particle states
(per unit volume) in zero field ,viz $\frac{2}{h^{3}}\int d^{3}p$
is replaced by $\sum_{\nu}(2-\delta_{\nu,0})\int_{-\infty}^{\infty}dp_{z}
\int_{-\frac{eBL_{x}}{2}}^{\frac{eBL_{x}}{2}}dp_{y}$ and the matrix 
elements have to be calculated by using the exact wave functions for 
electrons and protons obtained by solving the Dirac equation in the 
presence of magnetic field [11-15] . We are now in a position to evaluate
the various quatities required for the calculation of bulk viscosity, 
viz, A, C  and $\lambda$. 
 
\end{section}

\begin{subsection}{Weak Magnetic Field}
\noindent For weak magnetic field several Landau levels are populated
and the matrix elements remain essentially unchanged [13-14] and one needs to 
account for the correct phase space factor. For non-relativistic 
degenerate nucleons the decay rate constant $\lambda$ is given by

\begin{eqnarray}
\lambda
&=& \frac{17}{480\pi}G_{F}^{2}\cos_{\theta_{c}}^{2} T^{4}		
	 (1+3g_{A}^{2})eB\bar m_{p}\bar m_{n}\theta(p_{F}(p)+p_{F}(e)-p_{F}(n))\nonumber \\
&&{}	\sum_{\nu=0}^{\nu_{max}}[2-\delta_{\nu,0}]
	\frac{1}{\sqrt{\mu_{e}^{2}-m_{e}^{2}-2\nu eB}}	
\end{eqnarray}

which reduces to well known expressions [16-17] in the limit
$B\rightarrow 0$.
Using equations (30),(31) and (32) the coefficients A and C given in equations
(17) and (18) become
\begin{eqnarray}
A
&=& [-\frac{(\bar\mu_{n}^{2}-\bar m_{n}^{2})}{3\bar\mu_{n}}+
	\frac{\sum_{\nu=0}^{\nu_{max}}(2-\delta_{\nu,0})\sqrt{\bar\mu_{p}^{2}-
	\bar m_{p}^{2}-2\nu eB}}{\sum_{\nu=0}^{\nu_{max}}(2-\delta_{\nu,0})
	\bar\mu_{p}
	(\bar\mu_{p}^{2}-\bar m_{p}^{2}-2\nu eB)^{-\frac{1}{2}}}\nonumber \\
&&{} \frac{\sum_{\nu=0}^{\nu_{max}}(2-\delta_{\nu,0})\sqrt{\mu_{e}^{2}-
	m_{e}^{2}-2\nu eB}}{\sum_{\nu=0}^{\nu_{max}}(2-\delta_{\nu,0})
	\mu_{e}(\mu_{e}^{2}-m_{e}^{2}-2\nu eB)^{-\frac{1}{2}}}]
	-\frac{1}{2}(\frac{g_{\rho N}}{m_{\rho}})^{2}(n_{p}-n_{n}) 
\end{eqnarray}

\begin{eqnarray}
C
&=& [\frac{\pi^{2}}{\bar\mu_{n}^{2} \sqrt{\bar\mu_{n}^{2}-\bar m_{n}^{2}}}
	+\frac{2\pi^{2}}{eB} \frac{1}{\sum_{\nu=0}^{\nu_{max}}
	(2-\delta_{\nu,0})\bar\mu_{p}
	{(\bar\mu_{p}^{2}-\bar m_{p}^{2}-2\nu eB)}^{-\frac{1}{2}}}\nonumber \\
&&{} \frac{2\pi^{2}}{eB} \frac{1}{\sum_{\nu=0}^{\nu_{max}}
	(2-\delta_{\nu,0})\mu_{e}
	{(\mu_{e}^{2}-m_{e}^{2}-2\nu eB)}^{-\frac{1}{2}}}]
	+(\frac{g_{\rho N}}{m_{\rho}})^{2}
\end{eqnarray}

\end{subsection}

\begin{subsection}{Strong Magnetic Field}
\noindent For strong magnetic field , the electron is forced into the lowest
Landau level corresponding to $\nu=0$ . This happens for a field 
$B>B_{c}$ , where $B_{c}$ is given by 
$eB_{c}=\frac{\mu_{e}^{2}-m_{e}^{2}}{2}$.
Because of the equality $n_{p}=n_{e}$ , dictated by charge neutrality,
it is then straight forwad to verify that for $B>B_{c}$ only the
lowest Landau level $\nu=0$ contributes for the protons as well.
This leads to considerable simplification in the expression for the 
decay rate. Using the exact wave functions for protons and electrons in the lowest Landau level and following [15] , the rate for the process (1)
can be calculated and we get
\begin{eqnarray}
\Gamma
&=& \frac{2G_{F}^{2}\cos_{\theta_{c}}^{2}}{(2\pi)^{5}}eB
	\frac{\bar m_{n} \bar m_{p}}{p_{F}(p)}\int dE_{p} dE_{e} dE_{n} 
	dE_{\nu} E_{\nu}^{2}\delta(E_{p}+E_{e}-E_{n}-E_{\nu})
	(1-f_{n})f_{p}f_{e}\nonumber  \\
&&{}	[4g_{A}^{2}+(g_{A}+g_{v})^{2}]	
	[e^{-\frac{(p_{F}^{2}(n)-4p_{F}^{2}(e))}{2eB}}
	\theta(p_{F}^{2}(n)-4p_{F}^{2}(e))+e^{-\frac{p_{F}^{2}(n)}{2eB}}
	\theta(p_{F}(n))]
\end{eqnarray}
Carrying out the energy integrals for degenerate matter , the decay constant
$\lambda$ is given by
\begin{eqnarray}
\lambda
&=& \frac{17}{960\pi}G_{F}^{2}\cos_{\theta_{c}}^{2} eB
	\frac{\bar m_{n} \bar m_{p}}{p_{F}(p)} T^{4}
	[4g_{A}^{2}+(g_{v}+g_{A})^{2}]\nonumber  \\
&&{} [e^{-\frac{(p_{F}^{2}(n)-4p_{F}^{2}(e))}{2eB}}
	\theta(p_{F}^{2}(n)-4p_{F}^{2}(e)) + e^{-\frac{p_{F}^{2}(n)}{2eB}}
	\theta(p_{F}(n))]
\end{eqnarray}

The constants A and C are obtained from equations (34) and (35)
by setting $\nu=0$.
Substituting A, C and $\lambda$  into the expression (20)
the bulk viscosity can now be calculated for a given baryon density and 
temperature.

\end{subsection}

\begin{section}{Results and Discussions}

\noindent In this paper we have  used the relativistic mean field 
approximation to describe nuclear matter at high densities. This approach
 gives  a good description of nuclear matter and neutron star properties
at high densities. We have used the values of the various couplings 
$ g_{\sigma N} $ , $ g_{\omega N} $ , $ g_{\rho N} $ , b and c
given in [10] namely 
\begin{eqnarray} 
\frac{g_{\sigma N}}{m_{\sigma}}=0.0152502 MeV^{-1}\nonumber \\
 \frac{g_{\omega N}}{m_{\omega}}=0.011  MeV^{-1} \nonumber \\
 \frac{g_{\rho N}}{m_{\rho}}=0.011  MeV^{-1}\nonumber  \\
 b=3.418 \times 10^{-3} \nonumber  \\
c=0.0146 \nonumber  \\
\end{eqnarray}
With this set of parameters, one obtains
$ n_{B}=0.153 ~ fm^{-3} $ ; binding energy $ (\frac{E}{n_{B}}-m_{n})
= -16.3~ MeV $ ;  charge symmetry  energy =$ 32.5~  MeV $  ; 
bulk modulous  =$ 300~ MeV$ 
and  Landau  mass  = $ 0.83 ~m_{N} $. 

We have computed the bulk viscosity for zero magnetic 
field, for low magnetic fields by summing over various Landau levels and 
finally for very high magnetic fields, where only the $\nu=0$ level 
contributes.
In figure 1 we show the variation of the coefficient of bulk viscosity
with baryon density for $eB=0$ and $eB=10^{4}$ $MeV^{2}$ for $T=1~ MeV$.
As can be seen from (33) and (36) the variation of $\zeta$ with T 
goes as $T^{4}$; so for other temperatures $\zeta$ can be obtained by scaling.
We, in our study, have divided the magnetic field into two regimes, the ``low''
magnetic field where a large number of Landau levels contribute and 
the ``strong'' magnetic field where only the lowest Landau 
level contributes. The critical field $B_{c}$  separates the two regimes  
depends crucially on the chemical potential of the electron/proton.
These in turn depend upon the number density $n_{B}$. As $n_{B}$ varies from 
$0.2~fm^{-3}$ to $2.0~fm^{-3}$, the range of  $n_{B}$ over which figure 1 
is plotted, $eB_{c}$ remains 
well above $10^{4}~ MeV^{2}$. Therefore we have used the weak field formula
for computing viscosity at this value of magnetic field.
At zero magnetic field the viscosity changes smoothly with $n_{B}$. As $n_{B}$
changes from $0.2~fm^{-3}$ to $2.0~fm^{-3}$ viscosity changes by nearly 
one order of magnitude. At $eB=10^{4}$
$MeV^{2}$, superimposed on a nearly smooth change is a rapid fluctuation in 
viscosity with $n_{B}$. As one can see, whenever a new Landau 
level appears, the reaction rate shows a sudden discontinous change which
is reflected in a rapid fluctuation in viscosity.
In figure 2, we show the variation of viscosity with $n_{B}$ for ``high''
magnetic fields. This is achieved by considering only very small number 
densities, viz, $n_{B}\leq 0.01~ fm^{-3}$. For such small $n_{B}$, even 
at magnetic fields of a few hundreds $MeV^{2}$, only the lowest Landau 
level contributes and therefore one can use the formulas obtained in the 
high magnetic field case. The direct URCA process is now no longer inhibited 
whereas for $B=0$ the process is not allowed by energy momentum conservation and the bulk viscosity does not get any contribution.
We find that the viscosity decreases very rapidly with the changes in 
number density. For example, for $eB=100~ MeV^{2}$, the viscosity
changes by 12 orders of magnitude as $n_{B}$ varies from $0.001~fm^{-3}$ to 
$0.004~fm^{-3}$. The variation in viscosity with 
$n_{B}$ is most drastic at lower magnetic fields; at higer magnetic fields the 
variation with $n_{B}$ becomes much less, so much so that for 
$eB=1000~ MeV^{2}$
the bulk viscosity remains almost constant and beyond that, it  infact  starts 
increasing with $n_{B}$.
In figure 3 we plot bulk viscosity against the magnetic field for two baryon 
densities,viz, $n_{B}=0.34~ fm^{-3}$ and $n_{B}=0.6~ fm^{-3}$. For magnetic 
fields upto 10,000 $MeV^{2}$, the weak field approximation is valid as 
pointed out earlier. Here we find, not surprisingly, that the viscosity is 
nearly independent of magnetic field. As the magnetic field increases
(beyond $10^{4}~ MeV^{2}$ for $n_{B}=0.34~fm^{-3}$ and $3\times~ 10^{4}~MeV^{2}$
for $n_{B}=0.6~fm^{-3}$) bulk viscosity begins to increase with magnetic 
field because now only  the lowest Landau level contributes and the 
magnetic field can be considered ``high''.
In figure 4, we show the variation of viscosity with magnetic field at low 
densities. We find that initially the viscosity increases very rapidly but then 
saturates to a nearly constant value as the magnetic field increases.
\noindent To conclude we find that in the presence of strong magnetic field, 
the triangular inequality is no longer required to be satisfied
for the URCA process to contribute to the viscosity. At very
low densities this requirement can be met for magnetic field strengths
of a few hundred $MeV^{2}$, and the bulk viscosity of the medium is finite
though small. This is to be contrasted with the situation in which
the magnetic field is zero and the URCA process does not contribute
to bulk viscosity upto a certain minimum density. However
at densities of relevance in neutron stars($n_{B}\sim 2-4~n_{0}$),
the effect of non-zero magnetic field will be felt only at extremely
high values of B($\geq 10^{4}\sim~10^{5}$). even though the
reaction rate is directly proportional to the magnetic field,
the viscosity depends in a rather complicated fashion on the magnetic field.

\end{section}

%\pagebreak

%Figure Caption:
%\vskip 0.5cm 

%Figure 1:Bulk Viscosity $\zeta$ in units of $gm.cm.s^{-1}$ as a function
%of Baryon Density $n_{B}$ in $fm^{-3}$ at $T=1~ MeV$ for magnetic field
%$eB=0$ and $eB=10^{4}~MeV^{2}$.
%\vskip 0.5cm

%Figure 2: Bulk Viscosity $\zeta$ as a function of Baryon Density for 
%different values of the magnetic field at $T=1~MeV$. The curves 
%correspond to $eB=100, 200, 400, 600, 1000~ and~ 5000~ MeV^{2}$
%as we go up. The units of $\zeta$, $n_{B}$ and B are same as in
%Fig. 1.
%\vskip 0.5cm

%Figure 3: Bulk Viscosity $\zeta$ as a function of magnetic field at 
%$T=1~MeV$. The lower and upper graphs are for 
%$n_{B}=0.34~fm^{-3}~~and ~~0.6~fm^{-3}$ respectively.
%\vskip 0.5cm

%Figure 4: Bulk Viscosity $\zeta$ as a function of magnetic field at
%$T=1~MeV$. The lower and the upper curves are for 
%$n_{e}=0.01~fm^{-3}~~and~~0.004~fm^{-3}$ respectively.

\pagebreak

\pagebreak

%\begin{figure}
%\epsfig{file=bv1.ps,width=10cm,height=5cm}
%\begin{center}
%Figure 1
%\end{center}
%\end{figure}
\begin{figure}
\vskip 2truecm
\epsfig{file=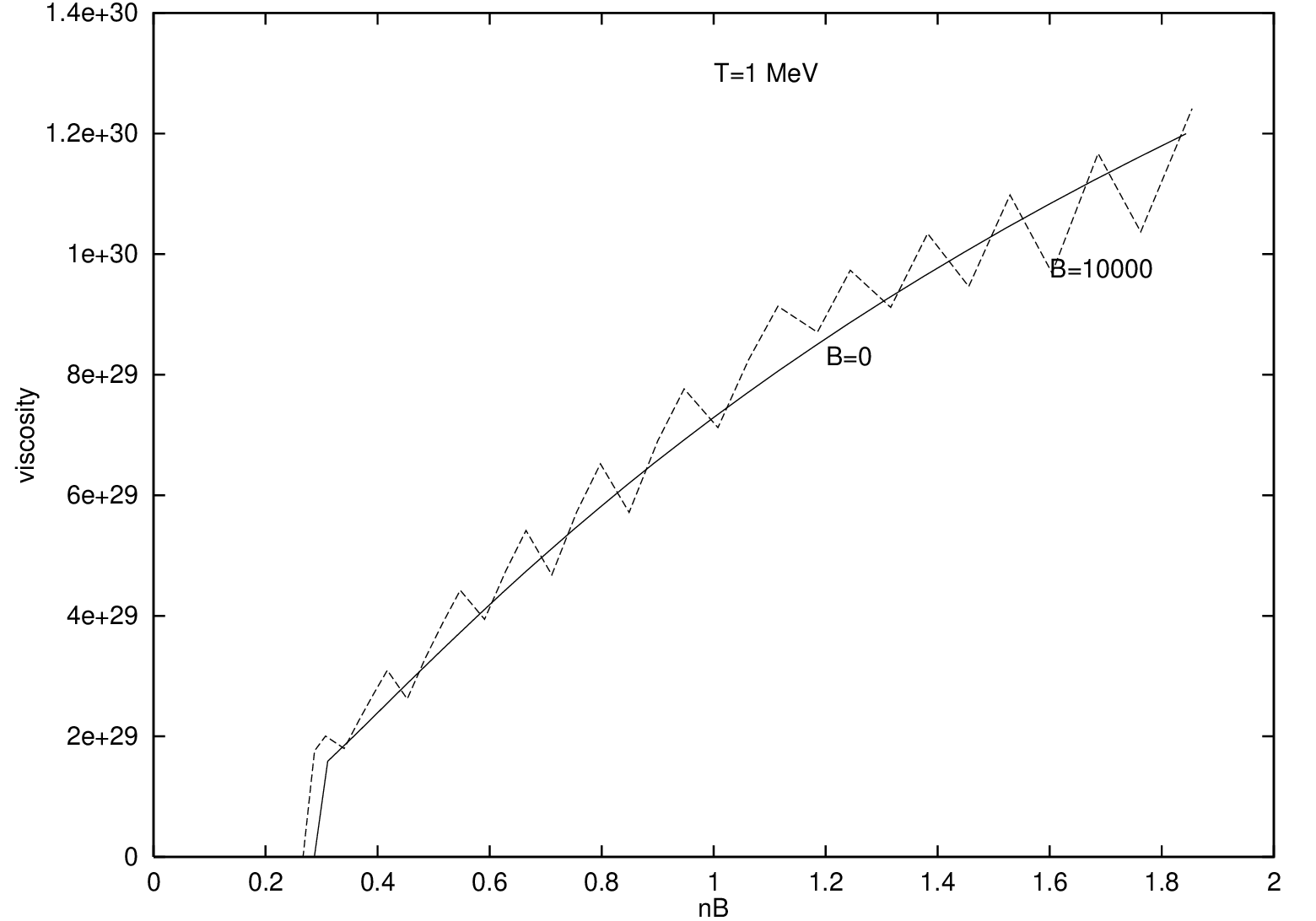,width=15cm,height=10cm}
%\begin{center}
Figure 1
%\end{center}
\caption{Bulk Viscosity  $\zeta$  in units of $gm.cm.s^{-1}$ as a function
of Baryon Density $n_{B}$ in $fm^{-3}$ at $T=1~ MeV$ for magnetic field
$eB=0$ and $eB=10^{4}~MeV^{2}$.}
\end{figure}
\vfill
\eject
\begin{figure}
\vskip 2truecm
\epsfig{file=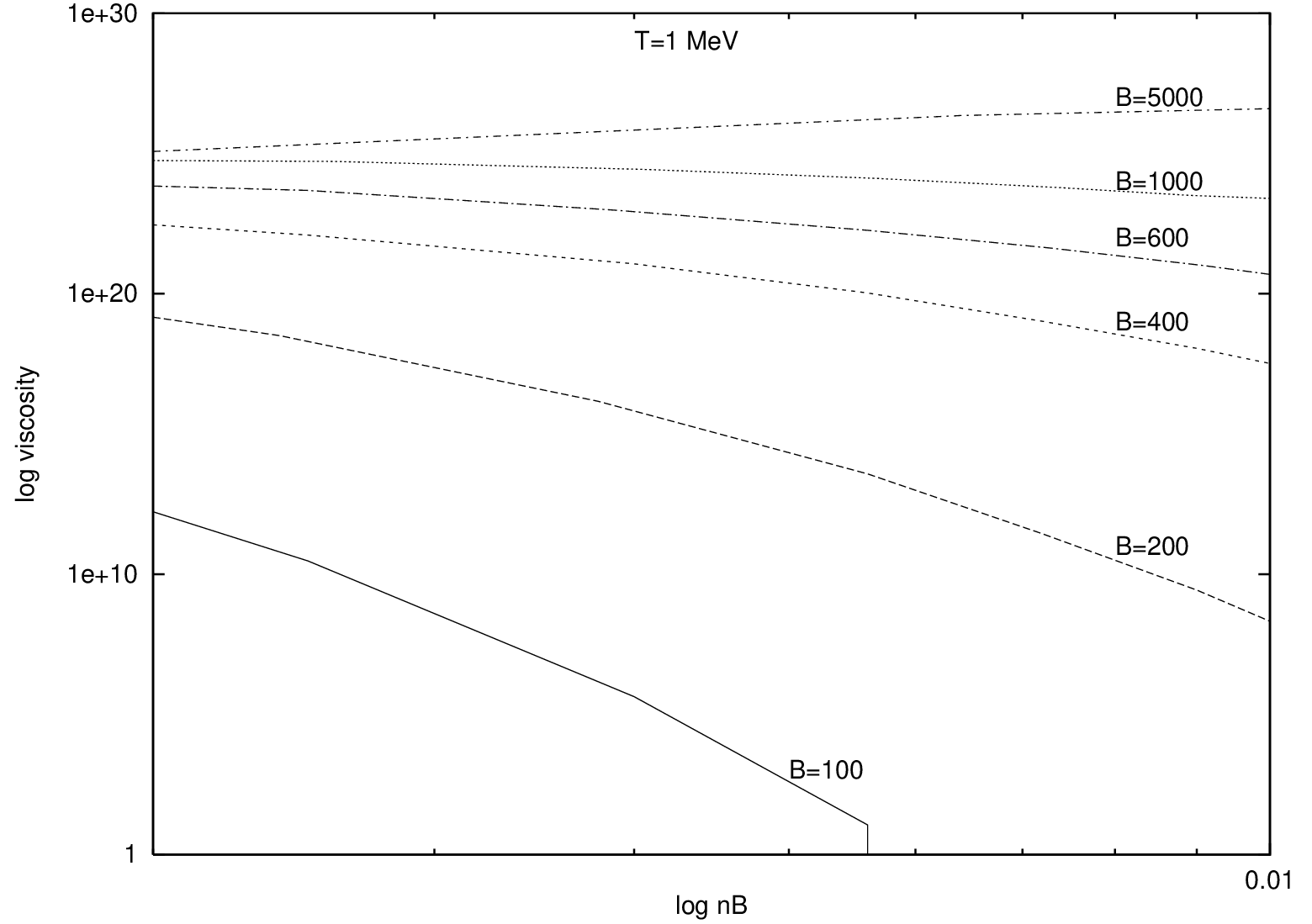,width=15cm,height=13cm}
%\begin{center}
Figure 2
%\end{centre}
\caption{Bulk Viscosity $\zeta$ as a function of Baryon Density $n_{B}$ 
for different values of the magnetic field at $T=1~MeV$. The units of 
$\zeta$, $n_{B}$ and B are same as in Fig. 1.}
\end{figure}
\begin{figure}
\vskip 2truecm
\epsfig{file=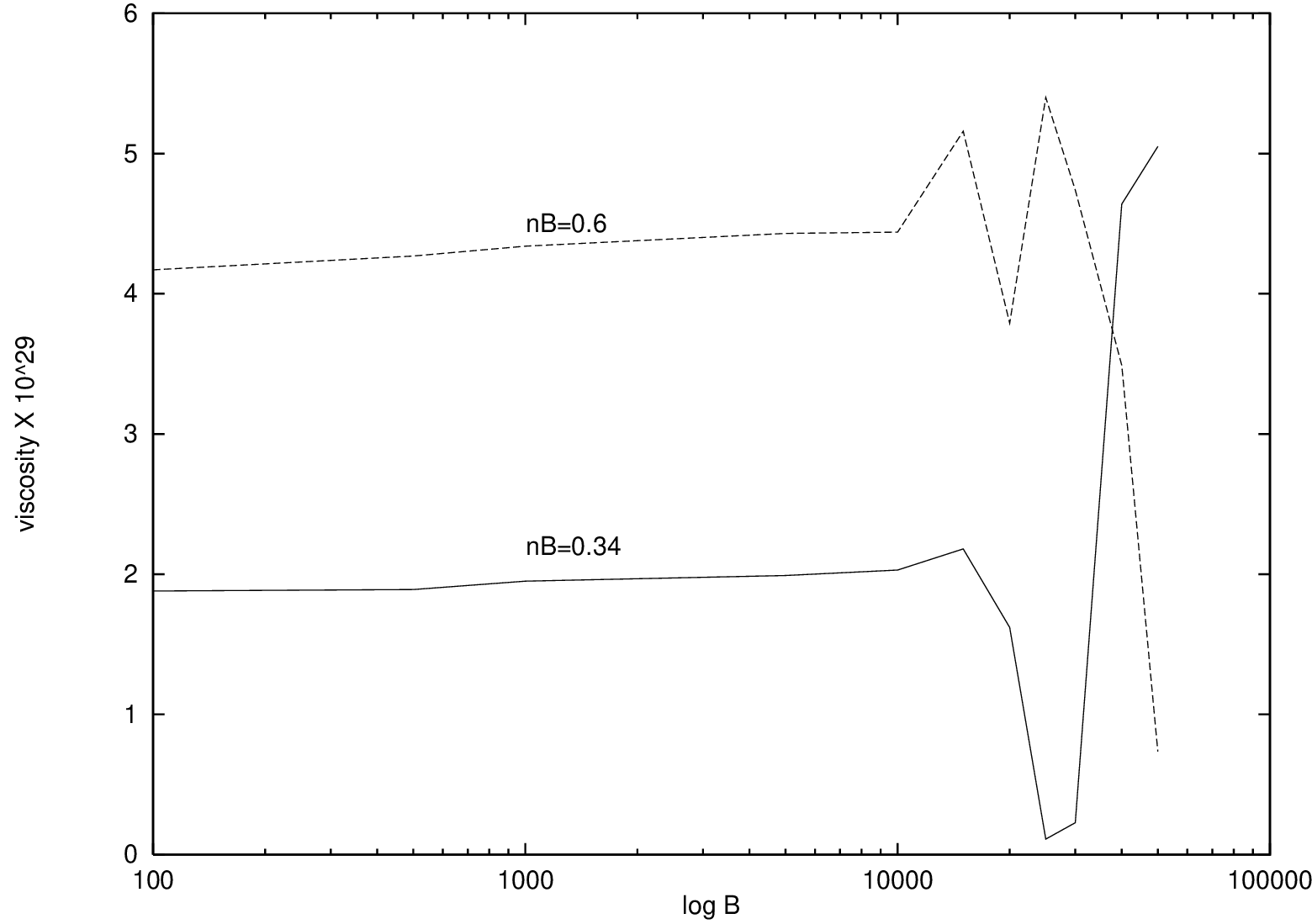,width=15cm,height=13cm}
%\begin{centre}
Figure 3
%\end{centre}
\caption{Bulk Viscosity $\zeta$ as a function of magnetic field at 
$T=1~MeV$ for different values of baryon density. The units of 
$\zeta$, $n_{B}$ and B are same as in Fig. 1.}
\end{figure}
\begin{figure}
\vskip 2truecm 
\epsfig{file=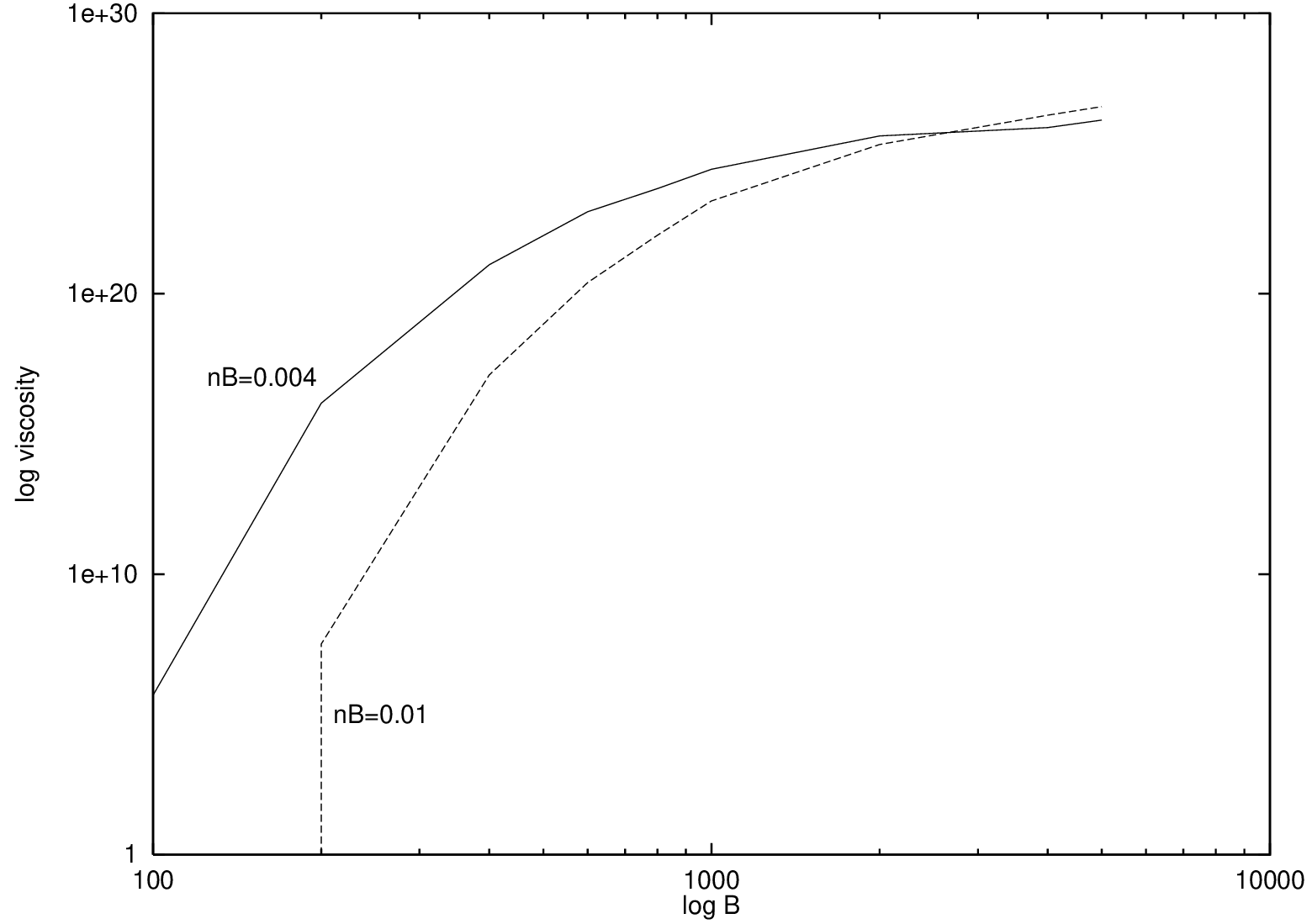,width=15cm,height=15cm}
%\begin{centre}
Figure 4
%\end{centre}
\caption{Bulk Viscosity $\zeta$ as a function of magnetic field at
$T=1~MeV$ for different values of baryon density. The units of 
$\zeta$, $n_{B}$ and B are same as in Fig. 1.}
\end{figure}

\end{document}